\documentclass{article}
\usepackage{spconf,amsmath,graphicx}
\usepackage{CJK}
\usepackage{algorithm}
\usepackage{algorithmicx}
\usepackage{algpseudocode}
\usepackage{amsmath}
\usepackage{multirow}
\usepackage{float}
\usepackage{bm}
\usepackage{setspace}

\title{A Novel Learnable Dictionary Encoding Layer for \\End-to-End Language Identification}
\name{Weicheng Cai$^{1,3}$, Zexin Cai$^{1}$, Xiang Zhang$^{3}$, Xiaoqi Wang$^{4}$ and Ming Li$^{1,2}$\sthanks{This research was funded in part by the National Natural Science Foundation of China (61401524,61773413), Natural Science Foundation of Guangzhou City (201707010363), Science and Technology Development Foundation of Guangdong Province (2017B090901045), National Key Research and Development Program (2016YFC0103905).}}

\address{$^1$School of Electronics and Information Technology, Sun Yat-sen University, Guangzhou, China\\
		$^2$Data Science Research Center, Duke Kunshan University, Kunshan, China\\
	$^3$Tencent Inc., Beijing, China\\
	$^4$Jiangsu Jinling Science and Technology Group Limited\\
	{\small \tt ml442@duke.edu}}

\begin{document}
	\ninept
	\maketitle
	\begin{abstract}

	A novel learnable dictionary encoding layer is proposed in this paper for  end-to-end language identification. It is inline with the conventional GMM i-vector approach both theoretically and practically.  We imitate the mechanism of traditional  GMM training and Supervector encoding procedure  on the top of CNN. The proposed layer can accumulate  high-order  statistics from variable-length input sequence and generate an utterance level fixed-dimensional vector representation. Unlike the conventional methods, our new approach provides an end-to-end learning framework, where the inherent dictionary are learned directly from the loss function. The dictionaries and the encoding representation for the classifier are learned jointly. The representation is orderless and therefore appropriate for language identification. We conducted a preliminary experiment on NIST LRE07 closed-set task,  and the results reveal that our proposed dictionary encoding layer achieves significant  error reduction comparing with the simple average pooling. 
	
		\end{abstract}
	\begin{keywords}
		language identification (LID), end-to-end,  dictionary encoding layer, GMM Supervector, variable length
	\end{keywords}
	\section{Introduction}
	\label{sec:intro}

	Language identification (LID) can be defined as a utterance level paralinguistic speech attribute classification task, in compared with automatic speech recognition, which is a ``sequence-to-sequence" tagging task. There is  no constraint on the lexicon words thus the training utterances  and testing segments may have completely different content \cite{Kinnunen2010An}. The goal, therefore, might be to find a robust and duration-invariant utterance level vector representation describing the distributions of local features. 

	In recent decades, in order to get the utterance level vector representation, dictionary learning procedure is widely used.  A dictionary, which contains several temporal orderless center components ( or units, words), can encode the variable-length input sequence into a single utterance level vector representation. Vector quantization (VQ) model, is one of the simplest text-independent dictionary models \cite{Kinnunen2010An}. It was introduced to speaker recognition in the 1980s \cite{Soong1985Report}. The average quantization distortion is aggregated from the frame-level residual towards to the K-means clustered codebook. The Gaussian Mixture Model (GMM) can be considered as an extension of the VQ model, in which the posterior assignments are soft \cite{Reynolds1995Robust,Reynolds2000Speaker}. Once we have a trained GMM, we can simply average the frame-level likelihood to generate the encoded utterance level likelihood score. Besides, we can move forward to accumulate the $0^{th}$ and $1^{st}$ order Baum-Welch statistics,  and encode them into a high dimensional GMM Supervector \cite{campbell2006support}.  VQ codebook and GMM are unsupervised and there is no exact physical meaning on its components. Another way to learn the dictionary is through phonetically-aware supervised training \cite{yun_icassp14,li2014interspeech}. In this method, a deep neural network (DNN) based acoustic model is trained. Each component in the dictionary represents a phoneme (or senone) physically, and the statistics is accumulated through senone posteriors, as is done in recently popular DNN i-vector approach \cite{Mclaren2015Advances,Richardson2015Deep,Snyder2016Time}. A phonotactic tokenizer can be considered as a dictionary doing hard assignments with top-1 score \cite{Kinnunen2010An}. Once we have a trained tokenizer, usually a bag-of-words (BoW)  or N-gram model is used to form the encoded representation \cite{Gelly2016A,Li2016Generalized}.

	These existing approaches have the advantage of accepting variable-length input and the encoded representation is in utterance level. However, when we move forward to modern the end-to-end learning pipeline, e.g. the neural network, especially for the fully-connected (FC) network,  it usually requires a fixed-length input. In order to feed into the network, as is done in \cite{lopez2014automatic,gonzalez2014automatic,Li2016Exploiting,Tkachenko2016Language}, the original input feature sequence has to be resized or cropped into multiple  small fixed-size segments in frame level. This might be theoretically and practically not ideal for recognizing language, speaker or other paralinguistic information due to the need of a time-invariant representation from the entire arbitrary and potentially long duration length.

	To deal with this issue, recently, in both \cite{1705.02304,Snyder2017Deep}, similar temporal average pooling(TAP) layer is adopted in their neural network architectures. With the merit of TAP layer, the neural network have the ability to train input segments with random duration. In testing stage, the whole speech segments with arbitrary duration can be fed into the neural network.

	Compared with the simple TAP, the conventional dictionary learning have the ability to learn a finer global histogram to demonstrate the feature distribution better, and it can accumulate high order statistics. In computer vision  community, especially in image scene classification, texture recognition, action recognition tasks, modern convolutional  neural network (CNN) usually bound to the conventional dictionary learning methods together to get a better encoding representation. For example,  NetVLAD \cite{Arandjelovic2016NetVLAD}, NetFV \cite{Simonyan2013Deep}, Bilinear Pooling \cite{Lin2016Bilinear}, and Deep TEN \cite{Zhang_2017_CVPR} are proposed and achieved great success. 
	

	This motivates us to implement the conventional GMM and Supervector mechanism into our end-to-end LID neural network. As the major contribution of this paper, we introduce a novel learnable dictionary encoding (LDE) layer, which  combines the entire dictionary learning and vector encoding pipeline into a single layer for end-to-end deep CNN.  The LDE layer imitates the mechanism of conventional GMM and GMM Supervector, but learned directly from the loss function. This representation is orderless which might be suitable for LID and many other test-independent paralinguistic speech attribute recognition tasks. The LDE layer acts as a smart pooling layer integrated on top of convolutional layers, accepting variable length inputs and providing output as an utterance level vector representation. By allowing variable-length inputs, the LDE layer makes the deep learning framework more flexible to train utterances with arbitrary duration. In these sense, it is inline with the classical  GMM i-vector \cite{dehak2010front} method both theoretically and practically.


\section{Methods}
\label{sec:methods}

\subsection{GMM Supervector}

In conventional GMM Supervector approach , all frames of  features in training dataset are grouped together to estimate a universal background model (UBM). Given a $C$ component GMM UBM model $\lambda$ with $\lambda_{c}=\{p_{c},\bm{\mu_{c}}, \bm{\Sigma_{c}}\},c=1,\cdots,C$ and an utterance with a $L$ frame feature sequence $\{\bm{x_{1}},\bm{x_{2}},\cdots,\bm{x_{L}}\}$, the $0^{th}$ and centered $1^{st}$ order Baum-Welch statistics on the UBM are calculated as follows:
\begin{equation}
N_{c}=\sum_{t=1}^{L}P(c|\bm{x_{t}},\lambda)
\label{eq_1}
\end{equation}
\begin{equation}
\bm{F_{c}}=\sum_{t=1}^{L}P(c|\bm{x_{t}},\lambda)  \bm{\cdot} \bm{r_{tc}} 
\label{eq_2}
\end{equation}

where $c=1,\cdots,C$ is the GMM component index and $P(c|\bm{x_{t}},\lambda)$ is the occupancy probability for $\bm{x_{t}}$ on $\lambda_{c}$. $\bm{r_{tc}}=\bm{x_{t}}-\bm{\mu_{c}}$ denotes as a residual between   $t^{th}$ frame feature and  the mean of the GMM's $c^{th}$ component.

The corresponding centered mean supervector $\tilde{\bm{F}}$ is generated by concatenating all the $\tilde{\bm{F_{c}}}$ together:
\begin{equation}
\tilde{\bm{F_{c}}}=\frac{\sum_{t=1}^{L}P(c|\bm{x_{t}},\lambda) \bm{\cdot} \bm{r_{tc}}}{\sum_{t=1}^{L}P(c|\bm{x_{t}},\lambda)}.
\label{eq_3}
\end{equation}

\begin{figure}[!htb]
	\centering
	\includegraphics[width=0.34\textwidth]{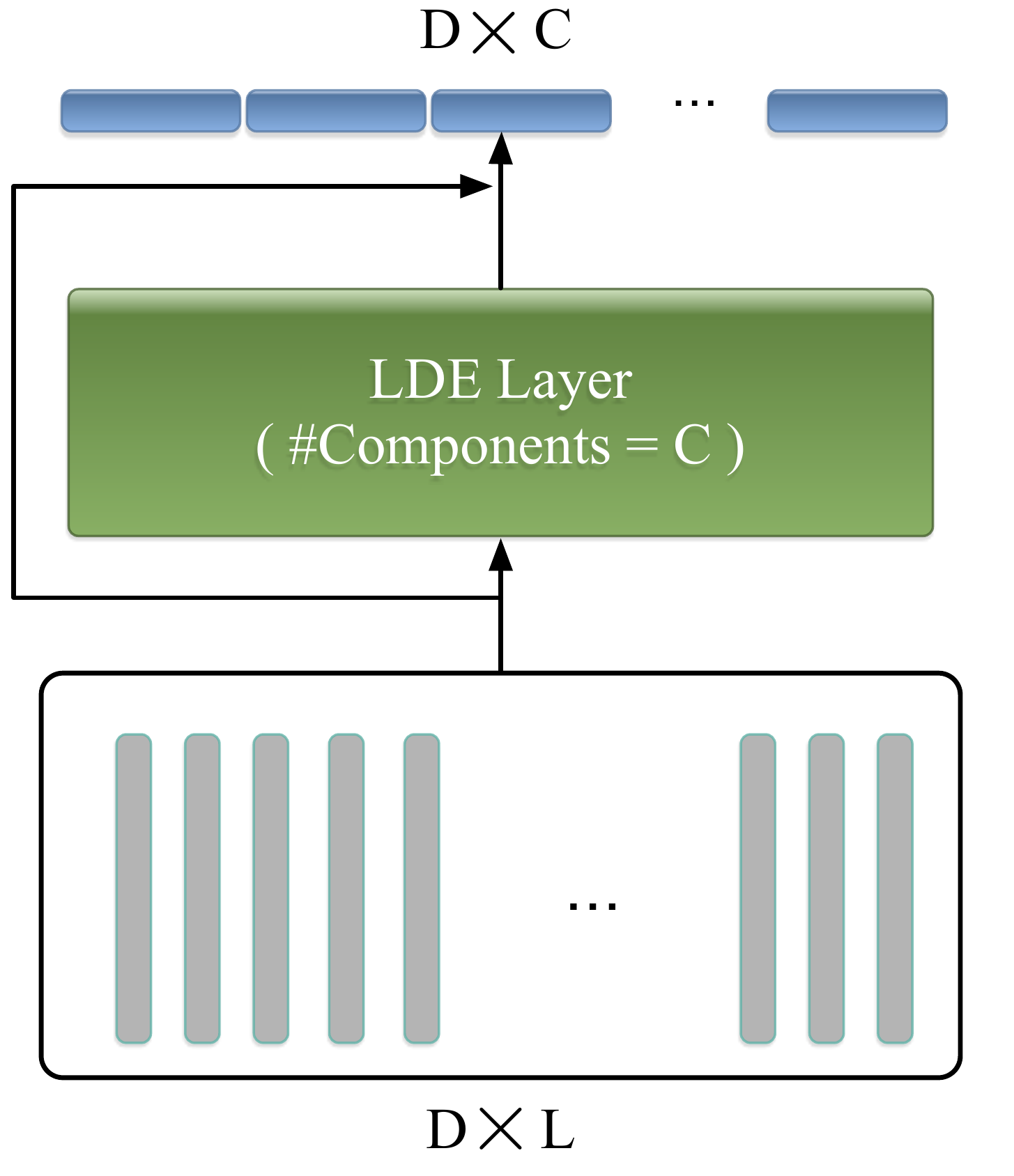}
	\caption{The input-out structure of LDE layer. It receives input feature sequence with variable length, produces an encoded utterance level vector with fixed dimension}\label{fig:lde_io}
\end{figure}
	
\subsection{LDE  layer}

Motivated by GMM Supervector encoding procedure, the proposed LDE layer has the similar input-output structure. As demonstrated in Fig. \ref{fig:lde_io},  given an input temporal ordered feature sequence with the shape $D \times L$ (where $D$ denotes the feature coefficients dimension, and $L$ denotes the temporal duration length), LDE layer aggregates them over time. More specifically, it transforms them into an utterance level temporal orderless $D \times C$ vector representation, which is independent of length $L$.

\begin{figure}[tb]
	\centering
	\includegraphics[width=0.495\textwidth]{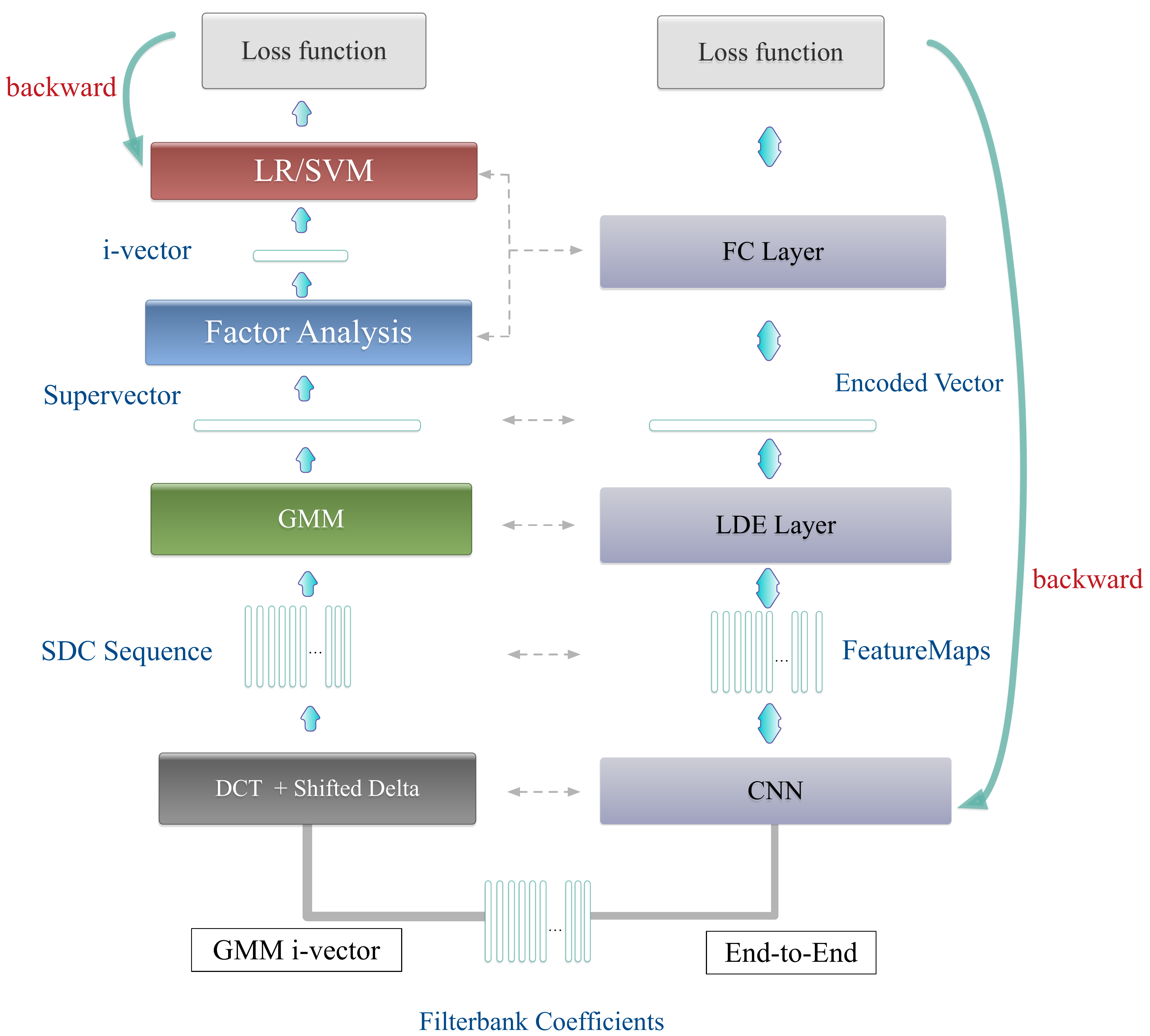}
	\caption{Comparison of GMM i-vector approach and end-to-end neural network with LDE layer}\label{fig:lde_e2e}
\end{figure}

Different from conventional approaches, we combine the dictionary learning and vector encoding into a single LDE layer on top of the front-end CNN, as shown in Fig. \ref{fig:lde_e2e}. The LDE layer simultaneously learns the encoding parameters along with an inherent dictionary in a fully supervised manner. The inherent dictionary is learned from the distribution of the descriptors by passing the gradient through assignment weights. During the training process, the updating of extracted convolutional features can also benefit from the encoding representations.

\begin{figure}[tb]
		\centering
		\includegraphics[width=0.39\textwidth]{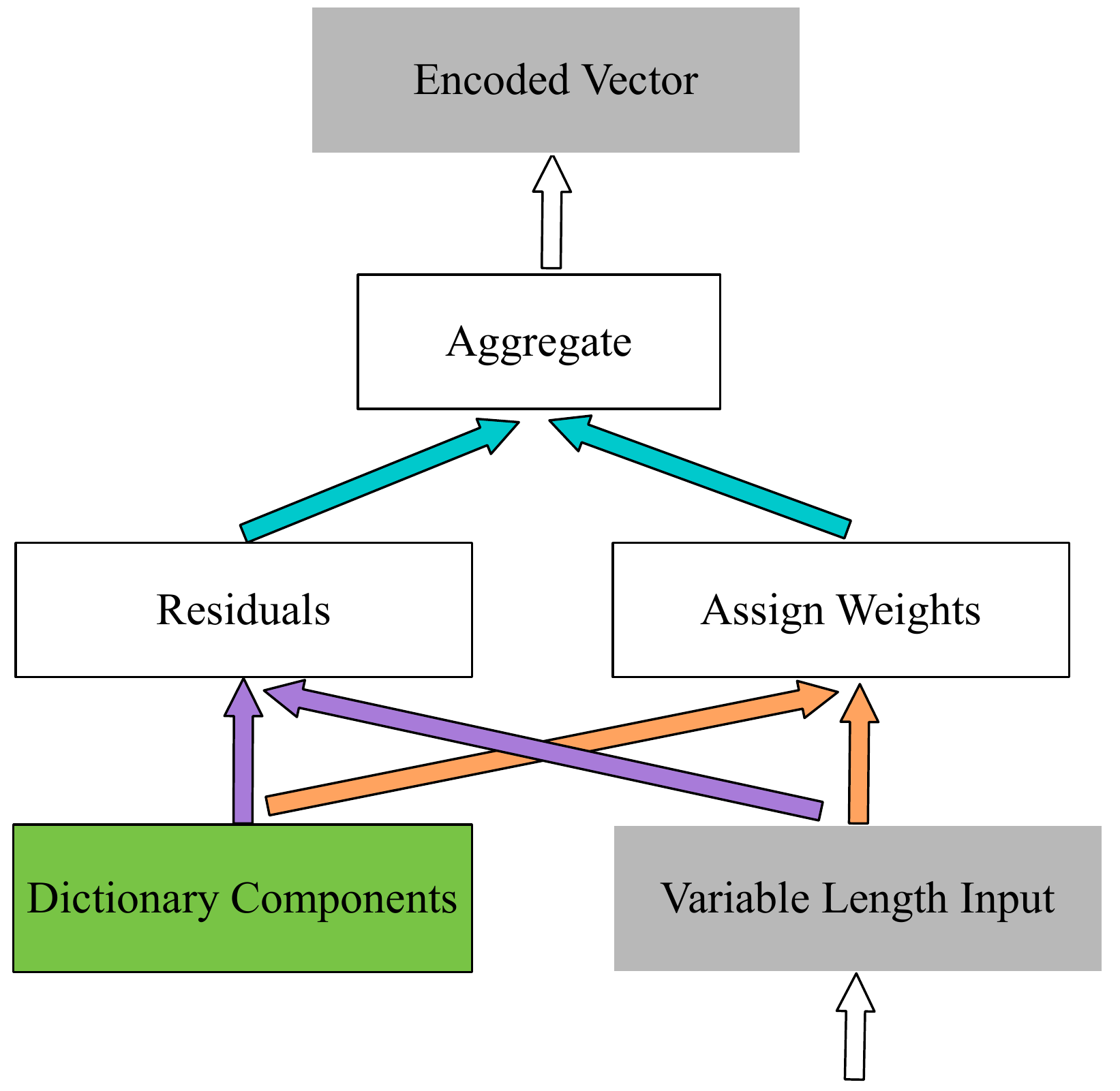}
		\caption{The forward diagram within the  LDE layer }\label{fig:forward_diagram}
\end{figure}

The LDE layer is a directed acyclic graph and all the components are differentiable $w.r.t$ the input $\bm{X}$ and the learnable  parameters. Therefore, the LDE layer can be trained end-to-end by standard stochastic gradient descent with backpropagation. Fig. \ref{fig:forward_diagram} illustrates the forward diagram of LDE layer. Here, we introduce two groups of learnable  parameters. One is the  dictionary component center, noted as $\bm{\mu} = \{\bm{\mu_1},  \cdots \bm{\mu_c} \}$. The other one is assigned weights, which is designed to  imitate the  $P(c|\bm{x_{t}},\lambda)$ , noted as $\bm{w}$.

 Consider  assigning weights from the features to the dictionary components. Hard-assignment provides a binary  weight for each feature $\bm{x_t}$, which corresponds to the nearest dictionary components. The $c^{th}$ element of the assigning vector is given by $w_{tc}=\delta(\Vert \bm{{r}_{tc}}\Vert ^2 = \min\{\Vert \bm{{r}_{t1}}\Vert, \cdots \Vert \bm{{r}_{tC}}\Vert\})$, where  $\delta$ is the indicator function (outputs 0 or 1).  Hard-assignment does not consider the dictionary component ambiguity and also makes the model nondifferentiable. Soft-weight assignment addresses this issue by assigning the feature to each dictionary component. The non-negative assigning weight is given by a softmax function,
 
\begin{equation}
w_{tc} = \frac{\exp(-\beta\Vert {\bm{r}_{tc}}\Vert ^2)}{\sum_{m=1}^{C}\exp(-\beta\Vert {\bm{r}_{tm}}\Vert ^2)}
\end{equation}

where $\beta$ is the smoothing factor for the assignment. Soft-assignment assumes that different clusters have equal scales. Inspired by GMM, we further allow the smoothing factor $s_c$ for each dictionary center $\bm{u_c}$ to be learnable:

\begin{equation}
w_{tc} = \frac{\exp(-s_c\Vert \bm{{r}_{tc}}\Vert ^2)}{\sum_{m=1}^{C}\exp(-s_i\Vert \bm{{r}_{tm}}\Vert ^2)}
\end{equation}

which provides a finer modeling of the feature distributions.

Given a set of $L$ frames feature  sequence $\{\bm{x_{1}},\bm{x_{2}},\cdots,\bm{x_{L}}\}$ and a learned dictionary center $\bm{\mu} = \{\bm{\mu_1},  \cdots \bm{\mu_c} \}$, each frame of feature $\bm{x_t}$ can be assigned with a weight $w_{tc}$ to each component $\bm{{\mu_c}}$ and the corresponding residual vector is denoted by $\bm{r_{tc}} = \bm{x_{t}}  -  \bm{u_{c}}$, where $t = 1,、\cdots L$ and $c = 1, \cdots C$. Given the assignments and the residual vector, similar to conventional GMM Supervector, the residual encoding model applies an aggregation operation for every dictionary component center $\bm{\mu_c}$:
\begin{equation}
\bm{e_{c}}=\sum_{t=1}^{L}e_{tc}=\frac{\sum_{t=1}^{L}(w_{tc} \bm{\cdot} \bm{r_{tc}})}{\sum_{t=1}^{L}w_{tc}}
\label{eq_2}
\end{equation}


It's complicated to compute the the explicit expression for the gradients of the loss $\ell$ with respect to the layer input $\bm{x_t}$. In order to facilitate the derivation， we simplified it as 
\begin{equation}
\bm{e_{c}}=\frac{\sum_{t=1}^{L}(w_{tc} \bm{\cdot} \bm{r_{tc}})}{L}
\label{eq_2}
\end{equation}

The LDE layer concatenates the aggregated
residual vectors with assigned weights. The resulted encoder outputs a fixed dimensional representation $\bm{E} = \{\bm{e_1} , \cdots \bm{e_C} \}$ (independent of the sequence length L). As is typical in conventional GMM Supervector/i-vector,  the resulting vectors are normalized using the length normalization \cite{garcia2011analysis}.

We implement the LDE layer similar as described in \cite{Zhang_2017_CVPR}, and more detail about the explicit expression for the gradients of the loss $\ell$ with respect to the layer input and the parameters can refer to \cite{Zhang_2017_CVPR}.

	\subsection{Relation to traditional  dictionary learning and TAP layer}
	Dictionary learning is usually learned from the distribution of the descriptors in an unsupervised manner. K-means learns the dictionary using hard-assignment grouping. GMM is a probabilistic version of K-means, which allows a finer modeling of the feature distributions. Each cluster is modeled by a Gaussian component with its own mean, variance and mixture weight. The LDE layer makes the inherent dictionary differentiable $w.r.t$ the loss function and learns the dictionary in a supervised manner. To see the relationship of the LDE to K- means, consider Fig. \ref{fig:forward_diagram} with omission of the residual vectors  and let smoothing factor $\beta\rightarrow\infty$. With these modifications, the LDE layer acts like K-means. The LDE layer can also be regarded as a simplified version of GMM, that allows different scaling (smoothing) of the clusters.

Letting $C$ = 1 and fixing $\bm{\mu}$ = 0, the LDE layer simplifies to TAP layer ($\bm{e} =  \frac{\sum_{t=1}^{L} \bm{x_t}}{L} )$ .

	\section{Experiments}
\label{sec:results}

\begin{table*} [tb]
	\caption{  Performance on the 2007 NIST LRE closed-set task }
	\centerline{
		\begin{tabular}{c c c c c c c c c c c }
			\hline
			{ System}& \multirow{2}{*}{ System Description}&\multirow{2}{*}{ Feature}&\multirow{2}{*}{Encoding Method}&\multicolumn{3}{c}{$C_{avg}(\%)$}&\multicolumn{3}{c}{$EER(\%)$}\\
			\cline{5-10}
			ID&&&&3s&10s&30s&3s&10s&30s\\
			\hline
			1&GMM i-vector&SDC  &GMM Supervector&20.46&8.29&3.02&17.71&7.00&2.27\\
			\hline
			2& CNN-TAP  &CNN FeatureMaps&TAP&9.98&3.24&1.73&11.28&5.76&3.96\\	
			3& CNN-LDE(C=16)  &CNN FeatureMaps&LDE&9.61&3.71&1.74&8.89&2.73&1.13\\	
			4& CNN-LDE(C=32)  &CNN FeatureMaps&LDE&8.70&2.94&1.41&8.12&2.45&0.98\\	
			5& \textbf{CNN-LDE(C=64)} &CNN FeatureMaps & LDE& \textbf{8.25}&\textbf{2.61}&\textbf{1.13}&\textbf{7.75}&\textbf{2.31}&\textbf{0.96}\\	
			6& CNN-LDE(C=128)  &CNN FeatureMaps&LDE&8.56&2.99&1.63&8.20&2.49&1.12\\
			7& CNN-LDE(C=256)  &CNN FeatureMaps&LDE&8.77&3.01&1.97&8.59&2.87&1.38\\	
			\hline
			8& Fusion ID2 + ID5&- & -&\textbf{6.98}&\textbf{2.33}&\textbf{0.91}&\textbf{6.09}&\textbf{2.26}&\textbf{0.87}\\
			\hline
	\end{tabular}}
	\label{table:lre07}
\end{table*}

\subsection{Data description}

We conducted experiments on 2007 NIST Language Recognition Evaluation(LRE). Our training corpus including  Callfriend datasets, LRE 2003, LRE 2005, SRE 2008 datasets, and development data for LRE07. The total training data is about 37000 utterances. 

The task of interest is the closed-set language detection. There are totally 14 target languages in testing corpus, which included 7530 utterances split among three nomial durations: 30, 10 and 3 seconds.

\subsection{GMM i-vector system}
For better result comparison, we built a referenced  GMM i-vector system based on Kaldi toolkit \cite{Povey_ASRU2011}. Raw audio is converted to 7-1-3-7 based 56 dimensional shifted delta coefficients (SDC) feature, and a frame-level energy-based voice activity detection (VAD) selects features corresponding to speech frames. All the utterances are split into  short segments no more than 120 seconds long. A 2048 components full covariance GMM UBM is trained, along with a 600 dimensional i-vector extractor, followed by length normalization and multi-class logistic regression.

\subsection{End-to-end system}
Audio is converted to 64-dimensional log mel-filterbank coefficients with a frame-length of 25 ms, mean-normalized over a sliding window of up to 3 seconds. The same VAD processing as in GMM i-vector baseline system is used here.  For improving the data loading efficiency, all the utterances are  split into short segments no more than 60s long , according to the VAD flags.

The receptive field size of a unit can be increased by stacking  more layers to make the network deeper or by sub-sampling. Modern deep CNN architectures like Residual Networks \cite{He2016Deep} use a combination of these techniques. Therefore, in order to get higher abstract representation better for utterances with long duration, we design a deep CNN based on the well-known ResNet-34 layer architecture, as is described in Table   \ref{resnetconfig}.

	\begin{table}[tb]\small

	\centering
	\caption{Our front-end CNN configuration}
	\label{resnetconfig}
	\resizebox{0.475\textwidth}{!}{
	\renewcommand\arraystretch{1.22}
	\begin{tabular}{|c|c|c|c|c|}
		\hline
		layer               & output size            & downsample      &  channels     &  blocks      \\ \hline
		conv1                     & $64$ $\times$ $L_{in}$                & False& 16       & -    \\ \hline
		res1                & $64$ $\times$  $L_{in}$               & False& 16& 3    \\ \hline
		res2                & 32 $\times$ $\frac{L_{in}}{2}$               & True& 32& 4    \\ \hline
		res3                & 16 $\times$ $\frac{L_{in}}{4}$              & True& 64& 6   \\ \hline
		res4                & 8 $\times$ $\frac{L_{in}}{8} $              & True& 128& 3    \\ \hline
		avgpool          & 1 $\times$ $\frac{L_{in}}{8} $ &-&128&- \\ \hline
		reshape         & 128$\times$  $L_{out}, L_{out}=\frac{L_{in}}{8}$ &-&-&- \\ \hline
		
	\end{tabular}}
	\end{table}

For CNN-TAP system, a simple average pooling layer followed with FC layer is built on top of the font-end CNN. For CNN-LDE system,  the average  pooling layer is replaced with a LDE layer.

\begin{figure}[tb]
	\centering
	\includegraphics[width=0.38\textwidth]{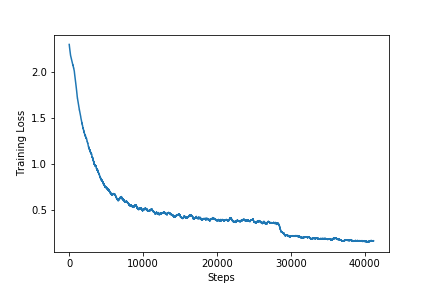}
	\caption{Loss duraing CNN-LDE training stage, smoothed with each 400 steps}\label{fig:loss}
\end{figure}

The network is trained using a cross entropy loss. The model is trained with a mini-batch, whose  size varies from 96 to 512 considering different model parameters. The network is trained for 90 epochs using stochastic gradient descent with momentum 0.9 and weight decay 1e-4. We start with a learning rate of 0.1 and divide it by 10 and 100 at 60th and 80th epoch. Because we have no separated validation set,  even though there might exist some model checkpoints can achieve better performance, we only use the model after the last step optimization. For each training step,  an integer $L$ within $\left[ 200 \textrm{,}  1000 \right]$  interval is randomly generated, and each data in the mini-batch is cropped or extended to $L$ frames. The training loss tendency of our end-to-end CNN-LDE neural network is demonstrated  in Fig. \ref{fig:loss}. It shows that our neural network with LDE layer is traninable and the loss can converge to a small value.

In testing stage, all the 3s, 10s, and 30s duration data is tested on the same model. Because the duration length is arbitrary, we feed the testing speech utterance to the trained neural network one by one.

In order to get the system fusion results of ID8 in Table \ref{table:lre07}, we randomly crop several additional training data corresponding to the separated 30s, 10s, 3s duration tasks. The score level system fusion weights are all trained on them.
 
\subsection{Evaluation}

Table \ref{table:lre07} shows the performance on the 2007 NIST LRE closed-set task.  
The performance is reported in average detection cost $C_{avg}$  and equal error rate (EER). Both CNN-TAP and CNN-LDE system achieve significant performance improvement comparing with conventional GMM i-vector system. 

For our purpose in exploring encoding method for end-to-end neural network,  we focus the comparison  on system ID2 and ID3-ID7.  The CNN-LDE system outperforms the CNN-TAP system with all different number of dictionary components. When the numbers of dictionary component increased from 16 to 64, the performance improved insistently. However, once dictionary component numbers are larger than 64, the performance decreased perhaps because of overfitting.

Comparing with CNN-TAP, the best CNN-LDE-64 system achieves significant performance improvement especially with regard to EER. Besides, their score level fusion result further improves the system performance significantly.

	\section{Conclusions}
	\label{sec:conclusion}
	In this paper, we imitate the GMM Supervector encoding procedure and introduce a LDE layer for end-to-end LID neural network. The LDE layer acts as a smart pooling layer integrated on top of convolutional layers, accepting arbitrary input lengths and providing output as a fixed-length representation. Unlike the simple TAP, it rely on a learnable dictionary and can accumulate more discriminative statistics. The experiment results show the superior and complementary of LDE comparing with TAP. 
	


	\newpage

	\bibliographystyle{IEEEbib}
	\bibliography{csl2}
	
\end{document}